# The ethylene – carbon dioxide complex and the double rotor model

A.R.W. McKellar,[1] and N. Moazzen-Ahmadi[2]

[1] *National Research Council of Canada, Ottawa, Ontario K1A 0R6, Canada*

[2] *Department of Physics and Astronomy, University of Calgary, 2500 University Drive North West, Calgary, Alberta T2N 1N4, Canada*

**Abstract**

The infrared spectrum of the weakly-bound $C_2H_4 – CO_2$ complex is investigated in the region of the $\nu_3$ fundamental band of $CO_2$ ($\approx 2350$ cm$^{-1}$), using a tunable OPO laser source to probe a pulsed supersonic slit jet expansion. The spacing of the various $K$-subbands in this perpendicular ($\Delta K = \pm 1$) spectrum is very irregular, and the pattern of irregularity is quite different from that observed previously in another $C_2H_4 – CO_2$ band by Bemish *et al*. [J. Chem. Phys. **103**, 7788 (1995)]. But by allowing for the different symmetry of the $\nu_3$ ($CO_2$) upper vibrational state, both results can be strikingly well explained using the 'double internal rotor' model as described by Bemish *et al*.

### 1. Introduction

The only previous high-resolution spectroscopic study of the weakly-bound complex $C_2H_4 - CO_2$ was reported by Bemish, Block, Pedersen and Miller,[1] who recorded its infrared spectrum in the region of the $\nu_9$ fundamental band of ethylene ($\approx 3100$ cm$^{-1}$). This spectrum was analyzed in terms of a number of perpendicular subbands such as $K_a = 1 \leftarrow 2, 0 \leftarrow 1, 1 \leftarrow 0, 2 \leftarrow 1$, etc., with $c$-type rotational selection rules. But the spacing of these subbands was found to deviate significantly from the value ($\approx 2A$, where $A$ is the rotational constant corresponding to the smallest principal moment of inertia) normally expected for a semi-rigid asymmetric rotor molecule. The nominal structure of the complex was determined to be "stacked parallel" as shown here in Fig. 1, with the $CO_2$ monomer lying in a symmetric position parallel to the plane of the $C_2H_4$ monomer. Bemish $et$ $al.$[1] were able to successfully explain the deviations noted in their spectrum by using a hindered internal rotation model in which the two monomers were partly free to rotate with respect to each other around the axis connecting their centers of mass. The barrier to this rotation was determined to be about 6.6 cm$^{-1}$.

In the present paper, we study the spectrum of $C_2H_4 - CO_2$ in the region of the $CO_2$ $\nu_3$ asymmetric stretch fundamental ($\approx 2350$ cm$^{-1}$). We observe similar perpendicular subbands (now with $b$-type selection rules) and similar large deviations of the subband spacing from the normal $2A$ value. But the pattern of our deviations is very different from that of Bemish $et$ $al.$[1]. It turns out that these differences can be remarkably well explained using the same hindered rotor model as long as we remember that only $odd$ values are allowed for the $CO_2$ rotational quantum number in the upper state of our transition ($CO_2$ $\nu_3$), as compared to only $even$ values in both the upper and lower states of their spectrum.

Some other reported spectra of weakly-bound complexes containing ethylene include ethylene dimer and trimer,[2-4] ethylene-rare gas,[5-7] ethylene-nitrous oxide,[8] and ethylene-hydrogen halide.[9,10]

## 2.  Results

### 2.1. Subband origins

Here we outline what we mean by 'subband origin' and how these origins were determined. Bemish *et al.*,[1] whose analysis we follow, did not explain exactly how they did this and did not report their subband origin values. We use a simple rigid near-prolate asymmetric rotor Hamiltonian whose matrix, after symmetrization, has diagonal terms given by

$$E(J, K) = AK^2 + \tfrac{1}{2}\,(B + C)[J(J+1) - K^2] \pm \tfrac{1}{4}\,(B - C)\,J(J+1),$$

where $A$, $B$, and $C$ are the usual rotational constants and the last ($\pm$) term applies only to $K = 1$ levels. Here and in the remainder of the paper, $K = K_a$ unless otherwise indicated. In addition to this diagonal term, the asymmetric rotor Hamiltonian contains off-diagonal terms involving ($B - C$) connecting basis states differing by 2 in $K$. This results in further shifts to all levels when the matrix is diagonalized.

Transition frequencies $\sigma$ are given by the difference between excited and ground state energies. Neglecting off-diagonal shifts, this is given by

$$\sigma = [\nu_0 + A'K'^2 - A''K''^2] + \text{terms in } B', C', B'', \text{ and } C''$$
$$= \quad \nu_{sb}(K', K'') \quad\quad + \text{terms in } B', C', B'', \text{ and } C'',$$

where $'$ and $''$ refer to the upper and lower states, respectively, and $\nu_0$ is the overall vibrational band origin. This defines the subband origin, $\nu_{sb}(K', K'')$. The off-diagonal shifts slightly complicate the determination of subband origins because they depend on the spacing of the different stacks of $K$-levels. For example, $K = 0$ and 2 energies depend on the energy difference between $K = 0$ and 2. But in the present case this $K$ spacing is irregular and not well represented by $AK^2$, so how can we do a rigid rotor fit without using $A$? To deal with this problem, we determined the subband origins in an iterative fashion, first starting with a fixed value of $A$, next fitting these preliminary subband origins with the internal rotor model described below, and then re-determining the subband origins using the internal rotor model for the $K$-stack energies. Fortunately the convergence of the iteration was rapid, because the off-diagonal shifts are relatively small.

### 2.2. Experimental spectrum

Spectra were recorded as described previously,[11-13] using a pulsed supersonic slit jet expansion probed by a rapid-scan optical parametric oscillator source. The gas expansion mixture contained about 0.02% carbon dioxide plus 0.06% ethylene in helium carrier gas with a backing pressure of about 17 atmospheres. Spectra were also recorded with a mixture containing only $CO_2$ and He to help verify which spectral features required $C_2H_4$. Wavenumber calibration was carried out by simultaneously recording signals from a fixed etalon and a reference gas cell containing room temperature $CO_2$. The effective spectral resolution was about 0.002 cm$^{-1}$ and the measurement accuracy for unblended lines was about 0.0002 cm$^{-1}$.

The central portion of the observed spectrum is shown in Fig. 2. Here we have 'clipped out' the known transitions of $CO_2$ monomer, $CO_2$-He dimer,[14] and $CO_2$-He$_2$ trimer[15,16] in order to show only features requiring $C_2H_4$. This is a $b$-type perpendicular band, with $\Delta K_a = \pm 1$, $\Delta K_c = \pm 1, \pm 3$, since the $CO_2$ $\nu_3$ transition moment lies along the $b$-axis of the complex (Fig. 1). In contrast, the $CO_2$-$C_2H_4$ band studied previously[1] was $c$-type, with $\Delta K_a = \pm 1$, $\Delta K_c = 0$, since the $C_2H_4$ $\nu_9$ transition moment lies along the complex $c$-axis. The simulated asymmetric rotor spectrum in Fig. 2 represents the observed spectrum fairly well, including only the $K = 1 \leftarrow 0$ and $0 \leftarrow 1$ subbands. But when we extended the simulation to include the higher-$K$ subbands in the rest of the band, it was not possible to achieve anything close to a satisfactory fit. This was similar to the problem encountered by Bemish et al.[1]

The entire spectrum is shown in Fig. 3. Here the strong and somewhat broad lines indicated with asterisks are probably due to larger clusters containing $CO_2$ and $C_2H_4$ (trimers, tetramers, etc.), but most of the remaining sharp lines could be successfully assigned to the $CO_2$-$C_2H_4$ dimer even though the asymmetric rotor simulation was poor.

A total of 108 transitions was assigned, with values of $J$ and $K$ ranging up to 8 and 5, respectively. They are given as Supplementary Information. We then analyzed these transitions using multiple rigid-rotor fits in the iterative fashion described above. This resulted in the subband origins shown as vertical lines at the top of Fig. 3, whose irregular spacing is immediately evident. Tables 1 and 2 list the origin values and rigid rotor parameters from the individual subband fits.

This irregular spacing is reminiscent of that encountered by Bemish et al.[1] (see their Fig. 2), but with an important difference. Their irregularities were roughly symmetric with respect to the overall band origin; for example, the spacing between the $K = 2 \leftarrow 3$ and $1 \leftarrow 2$ subbands was similar to that between the $2 \leftarrow 1$ and $3 \leftarrow 2$ subbands. However, our irregularities (Fig. 3) are distinctly not symmetric around the band origin; for example, the $K = 2 \leftarrow 3$ to $1 \leftarrow 2$ spacing is much smaller than the $2 \leftarrow 1$ to $3 \leftarrow 2$ spacing. This asymmetry is closely related to the fact that our upper state ($CO_2$ $\nu_3$) has a different symmetry than the ground state. As shown below, both the previous and current results can be very well explained using the same internal rotor model proposed in Ref. 1.

Significant broadening was observed for some of the lines in our spectrum, which we attribute to shortened lifetimes in the excited state, that is, to predissociation. But the exact widths were difficult to measure because of the crowded nature of the spectrum. Broadening was most noticeable for the upper asymmetry doubling components of $K' = 1$ (that is, $J_{KaKc} = 1_{10}, 2_{11}$, etc.). These levels are involved in the strong but messy $K = 1 \leftarrow 0$ $Q$-branch around 2348.0 cm$^{-1}$. We estimated widths of roughly 0.002 cm$^{-1}$, or greater, for $J = 1, 2$, and 4, and 0.006 cm$^{-1}$, or greater, for $J = 3$. The lower components of $K' = 1$, involved in the $K = 1 \leftarrow 0$ $P$- and $R$-branches, were only slightly broadened ($\approx$0.001 cm$^{-1}$), at least for $J = 1$ to 5. Similar broadening of $K' = 1$ transitions was of course also observed in the $K = 1 \leftarrow 2$ subband. Levels with $K' = 0, 3$, and 5 showed little or no obvious broadening. Levels with $K' = 2$ had widths of roughly 0.003 cm$^{-1}$. The level with $K' = 4$, $J = 4$ was significantly broadened ($\approx$0.006 cm$^{-1}$), but the higher-$J$ $K' = 4$ levels were sharper ($\leq$0.002 cm$^{-1}$).

### 2.3. Internal rotation model

In the internal rotation model described by Bemish et al.,[1] the $CO_2$ and $C_2H_4$ monomers in Fig. 1 are free to rotate around the axis connecting their centers of mass, hindered only by a simple sinusoidal potential energy barrier $V(\theta) = a \cos(2\theta)$, where $\theta$ is the angle between the $CO_2$ axis and the C-C axis of $C_2H_4$. The Hamiltonian matrix resulting from this model has diagonal elements given by $B(CO_2)M^2(CO_2) + C(C_2H_4)M^2(C_2H_4)$, where $B$ and $C$ are monomer rotational constants and $M$ represents the quantum numbers for rotation around the axis. In the low-barrier limit ($a = 0$), the resulting energy is the sum of two independent one-dimensional rotors, and the $K_a$-value is the sum of $M(CO_2)$ and $M(C_2H_4)$. The off-diagonal elements of the matrix have a value of $a/2$ and connect states of $M(CO_2)$ or $M(C_2H_4)$ differing by 2. The basis set in our calculations included values of -14 to +14 for $M(CO_2)$, and -8 to +8 for $M(C_2H_4)$.

For the ground vibrational state of $CO_2$-$C_2H_4$, only even values of $M(CO_2)$ are allowed by nuclear spin statistics. The resulting correlation diagram showing energies as a function of $a$, illustrated here in Fig. 4, is very similar to that in Fig. 6 of Ref. 1. Note that energies for negative $M$-values are identical to those shown here for positive $M$. However, only odd values of $M(CO_2)$ are allowed in the upper state of our observed $CO_2$-$C_2H_4$ infrared band, corresponding to the $\nu_3$ vibration of $CO_2$. The correlation diagram for this case, shown in Fig. 5, is quite different, but it should be the one applicable to our upper state levels. The difference between the two cases is further illustrated in Fig. 6, which shows how the $K_a$ stack origins are shifted relative to a rigid rotor for even and odd $M(CO_2)$.

We used the internal rotation model to fit the observed subband origins. As mentioned above, the analysis was iterative, in that the origins were first approximately determined, then fitted with the internal rotation model, and then refined using the model results for the $K$ stack energies to make a series of new rigid rotor fits to each individual subband. Correct assignment of the subbands involving $K = 4$ and 5 was not obvious at first (because of irregular spacing) and was greatly aided by the preliminary model fits. The final self-consistent results of the analysis are given in Table 1 (calculated subband origins), Table 2 (rigid rotor rotational constants), and Table 3 (fitted band origin and internal rotation parameters). In the final fit, we gained a significant improvement by allowing the monomer rotational constants, $B(CO_2)$ and $C(C_2H_4)$ to vary, rather than fixing them at the free monomer values. This resulted in slightly increased values for these parameters, +0.9% for $B(CO_2)$ and +2.9% for $C(C_2H_4)$. We can rationalize these increases by considering the rocking motion of each monomer relative to the $a$-axis (Fig. 1), which reduces their projection on the $a$-axis, thus reducing the moment of inertia and increasing the effective rotational constant. The same effect occurs, for example, in the T-shaped $CO_2$-Ar complex, whose $A$-value is about 1.8% larger than $B(CO_2)$, which it would equal if the structure were rigid.[17] So it is quite reasonable that the rotational constants in our internal rotation model should be slightly larger than their free monomer values.

### 3. Discussion and conclusions

The hindered internal rotor analysis of the observed $CO_2$-$C_2H_4$ subband origins (Table 1), which involves 10 observed values and 4 adjustable parameters, fits remarkably well. It gives us a value for the overall band origin, $\nu_0 = 2347.683$ cm$^{-1}$, which represents a vibrational shift of -1.460 cm$^{-1}$ relative to the free $CO_2$ monomer origin. But the most interesting result is perhaps the height of the barrier to internal rotation, $2a$. Our value of $2a = 6.96(5)$ cm$^{-1}$ is about 5% larger than the value of 6.6 cm$^{-1}$ obtained by Bemish et al.,[1] which we consider to be good agreement. As for the reason for the 5% difference, this is difficult to investigate because details of the analysis (e.g. subband origins) are not given in Ref. 1. We can, however, say that the difference is not due to the fact that we allowed $B(CO_2)$ and $C(C_2H_4)$ to vary in the fit, because this did not significantly affect the resulting value of $2a$.

Table 2 compares our ground state rigid rotor rotational parameters with those of Bemish et al.[1] for $K = 0$, 1, and 2. The agreement is very good for all the $(B + C)/2$ values, as well as for $(B - C)$ of $K = 1$. The agreement is not so good for $(B - C)$ of $K = 0$ and 2, but this is perhaps understandable because $(B - C)$ is less directly determined in these cases because it is off-diagonal and thus more dependent on exactly how the rigid rotor analysis was done. Table 2 also shows a tendency for a slight decrease (of the order of 0.0001 cm$^{-1}$) in the value of $(B + C)/2$ for our excited state relative to the ground state, which is quite reasonable.

To date, the microwave pure rotational spectrum of $CO_2$-$C_2H_4$ has not been observed. This spectrum will be relatively weak because it depends on a small induced dipole moment, but should still be detectable thanks to the high experimental sensitivity possible in the microwave region. It will be have $a$-type selection rules, and will thus tend to give precise information on the ground state $B$ and $C$ rotational parameters, but not on $A$. Unfortunately, therefore, microwave spectra would probably not tell us any more about the most interesting aspect of the rotational energy level structure, namely the spacing between levels with different $K_a$-values.

In conclusion, the vibration-rotation band of the $CO_2$-$C_2H_4$ complex corresponding to the $\nu_3$ fundamental of $CO_2$ has been observed and analyzed. The spectrum was detected in a pulsed supersonic slit jet expansion as probed using a tunable OPO laser source. The band has $b$-type rotational selection rules, $\Delta K_a = \pm 1$, and distinct perpendicular subbands with $K_a = 1 \leftarrow 0$, $0 \leftarrow 1$, $2 \leftarrow 1$, etc. The spacing of these subbands was found to be irregular, and the pattern of irregularity was very different from that observed in the only previous observation of $CO_2$-$C_2H_4$, made by Bemish et al.[1] in the $C_2H_4$ $\nu_9$ fundamental region. The irregular spacings in both cases can be remarkably well explained by a simple hindered internal rotation model,[1] as long as the different symmetry of the present upper state ($CO_2$ $\nu_3$) is taken into account. We obtain a value of $2a = 6.96(5)$ cm$^{-1}$ for the height of the barrier to internal rotation, slightly larger than the value of 6.6 cm$^{-1}$ determined previously.[1]

**Supplementary Material**

Supplementary Material includes a table giving observed and fitted line positions for $CO_2$-$C_2H_4$.

**Acknowledgements**

The financial support of the Natural Sciences and Engineering Research Council of Canada is gratefully acknowledged. We thank A.R. Barclay for help with the experiment.

Table 1. Observed and calculated subband origins for $CO_2$-$C_2H_4$ (in cm$^{-1}$). [a]

| $K' \leftarrow K''$ | $\nu_{sb}(K', K'')$ Observed | $\nu_{sb}(K', K'')$ Calculated | Obs - Calc |
|---|---|---|---|
| $4 \leftarrow 5$ | 2345.1137 | 2345.1080 | 0.0057 |
| $3 \leftarrow 4$ | 2345.8307 | 2345.8320 | -0.0013 |
| $2 \leftarrow 3$ | 2346.4010 | 2346.4030 | -0.0020 |
| $1 \leftarrow 2$ | 2346.7377 | 2346.7399 | -0.0022 |
| $0 \leftarrow 1$ | 2347.4827 | 2347.4831 | -0.0004 |
| $1 \leftarrow 0$ | 2348.0034 | 2348.0040 | -0.0006 |
| $2 \leftarrow 1$ | 2348.3448 | 2348.3495 | -0.0047 |
| $3 \leftarrow 2$ | 2349.1051 | 2349.1046 | 0.0005 |
| $4 \leftarrow 3$ | 2349.5825 | 2349.5785 | 0.0040 |
| $5 \leftarrow 4$ | 2349.9379 | 2349.9369 | 0.0010 |

[a] Meaningful uncertainties for the observe origins are not available due to the iterative nature of the analysis, and the last quoted digits (0.0001 cm$^{-1}$) are not necessarily significant. Calculated values are given by the hindered rotor model using the parameters listed in Table 3.

Table 2. Rigid rotor rotational constants for $K$ states of $CO_2$-$C_2H_4$ (in cm$^{-1}$). [a]

| $K$ | Upper state this work | | Ground state this work | | Ground state Bemish et al.[1] | |
| --- | --- | --- | --- | --- | --- | --- |
| | $(B' + C')/2$ | $(B' - C')$ | $(B'' + C'')/2$ | $(B'' - C'')$ | $(B'' + C'')/2$ | $(B'' - C'')$ |
| 0 | 0.07464 | 0.01669 | 0.07483 | 0.01706 | 0.0748 | 0.0117 |
| 1 | 0.07457 | 0.01684 | 0.07475 | 0.01671 | 0.0747 | 0.0167 |
| 2 | 0.07472 | 0.01650 | 0.07477 | 0.01716 | 0.0749 | 0.0146 |
| 3 | 0.07471 | [0.01650] | 0.07478 | [0.01670] | | |
| 4 | 0.07456 | [0.01650] | 0.07453 | [0.01670] | | |
| 5 | 0.07431 | [0.01650] | 0.07475 | [0.01670] | | |

[a] Meaningful uncertainties for the observe origins are not available due to the iterative nature of the analysis. Quantities in square brackets were fixed at the indicated values.

Table 3. Parameters resulting from the internal rotor model fit to the subband origins of $CO_2$-$C_2H_4$ (in $cm^{-1}$). [a]

| Parameter | Value | Value for free $CO_2$ or $C_2H_4$ monomer |
|---|---|---|
| $\nu_0$ | 2347.6829(11) | |
| $2a$ | 6.962(54) | |
| $B''_{eff}$ ($CO_2$) | 0.39363(85) | 0.390219 |
| $C_{eff}$($C_2H_4$) | 0.8519(39) | 0.828042 |

[a] Uncertainties in parentheses are $1\sigma$ in units of the last quoted decimal. $B'_{eff}$($CO_2$) was constrained to equal ($B''_{eff}$($CO_2$) − 0.00308) $cm^{-1}$, the same as in the monomer. The ground and excited state values of the barrier height, $2a$, and ethylene effective $C$-value, $C_{eff}$($C_2H_4$), were constrained to be equal.

**Figure Captions**

Fig. 1. Proposed equilibrium structure of $CO_2$-$C_2H_4$. The double-rotor model involves internal rotation of one monomer with respect to the other around the $a$-inertial axis. The center of mass separation of $CO_2$ and $C_2H_4$ is about 3.43 Å.[1]

Fig. 2. Central portion of the spectrum of $CO_2$-$C_2H_4$ in the region of the $\nu_3$ fundamental band of $CO_2$. Known lines due to $CO_2$, $CO_2$-He, and $CO_2$-$He_2$ have been 'clipped out', resulting in some visible gaps. The simulated spectrum, including only $K = 1 \leftarrow 0$ and $0 \leftarrow 1$ subbands, represents this part of the observed spectrum fairly well, but fails when extended to the higher-$K$ subbands on either side of this region.

Fig. 3. Observed spectrum of $CO_2$-$C_2H_4$ in the region of the $\nu_3$ fundamental band of $CO_2$. Vertical lines indicate the irregularly-spaced subband origins $(K' - K'')$ obtained by analyzing the spectrum. The blue arrow shows the constant subband spacing that would be expected in a normal rigid molecule. The features marked with asterisks are believed to be due to larger clusters (trimer, tetramer, …) containing $CO_2$ and $C_2H_4$.

Fig. 4. Hindered internal rotor correlation diagram for even values of $M(CO_2)$, suitable for the ground state of $CO_2$-$C_2H_4$. Energies are expressed relative to the lowest level, $M(CO_2) = M(C_2H_4) = K_a = 0$. Levels with even or odd values of $M(C_2H_4)$ are shown in black or red, respectively. The blue vertical line indicates the barrier height value determined by fitting the current spectrum. Parameters $B(CO_2)$ and $C(C_2H_4)$ have the fitted values from Table 3. The left-hand side ($2a = 0$) corresponds to free (one-dimensional) internal rotation. The right-hand side (large value of $2a$) approaches the rigid rotor case. The $K$ labels on the right-hand side are belong to the ground bending state, $v(bend) = 0$. The unlabeled levels whose energies increase as a function of $2a$ belong to excited bending states (see also Fig.

6 of Bemish et al.). For example, the upper of the two levels arising from $M(CO_2) = 2$, $M(C_2H_4) = 1$ is the $K = 1$ level of the v(bend) = 1 state.

Fig. 5.   Hindered internal rotor correlation diagram for odd values of $M(CO_2)$, suitable for the excited state of $CO_2$-$C_2H_4$ with $v_3(CO_2) = 1$. Energies are expressed relative to the lowest level in Fig. 4, $M(CO_2) = M(C_2H_4) = K_a = 0$. Levels with even or odd values of $M(C_2H_4)$ are shown in black or red, respectively. The left-hand side ($2a = 0$) corresponds to free (one-dimensional) internal rotation. The right-hand limit (large value of $2a$) approaches the rigid rotor case.

Fig. 6.   Energy level diagram for $K_a$ states of $CO_2$-$C_2H_4$ comparing the rigid rotor limit with the hindered internal rotor model (with $2a = 6.962$ cm$^{-1}$) for the case of even $M(CO_2)$ (the ground vibrational state) and odd $M(CO_2)$ (the $v_3(CO_2) = 1$ excited state). The large difference between even and odd $M(CO_2)$ explains why the current spectrum looks very different from that of Bemish et al.[1], and why the same model can explain both very well.

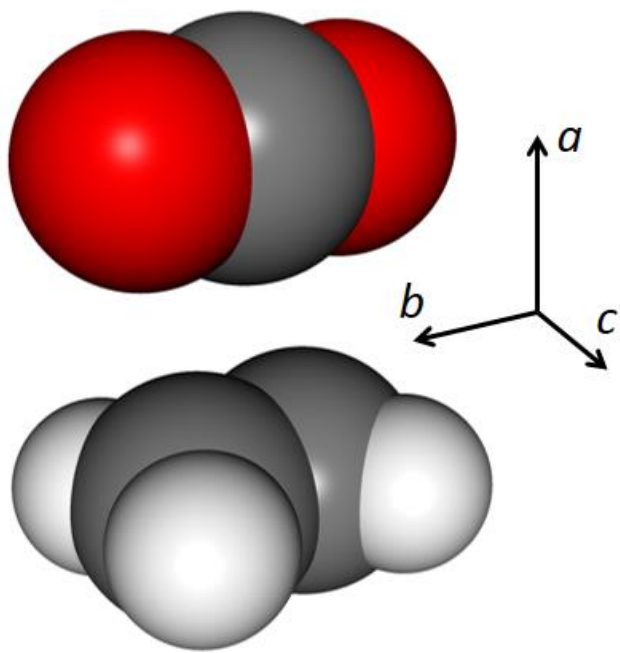

Fig. 1

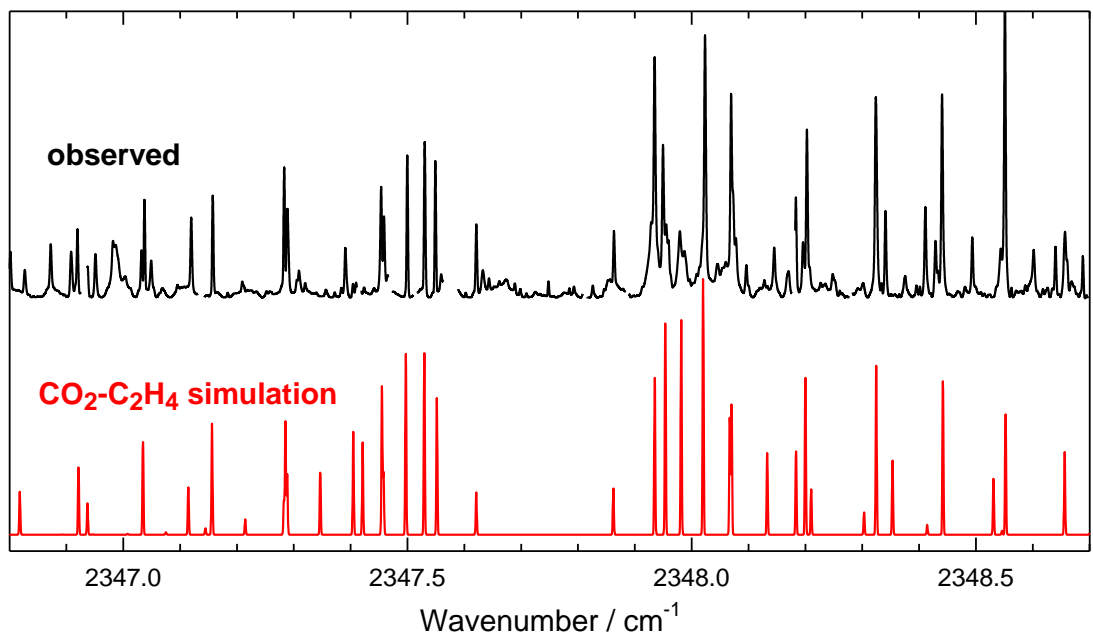

observed

CO₂-C₂H₄ simulation

Fig. 2.

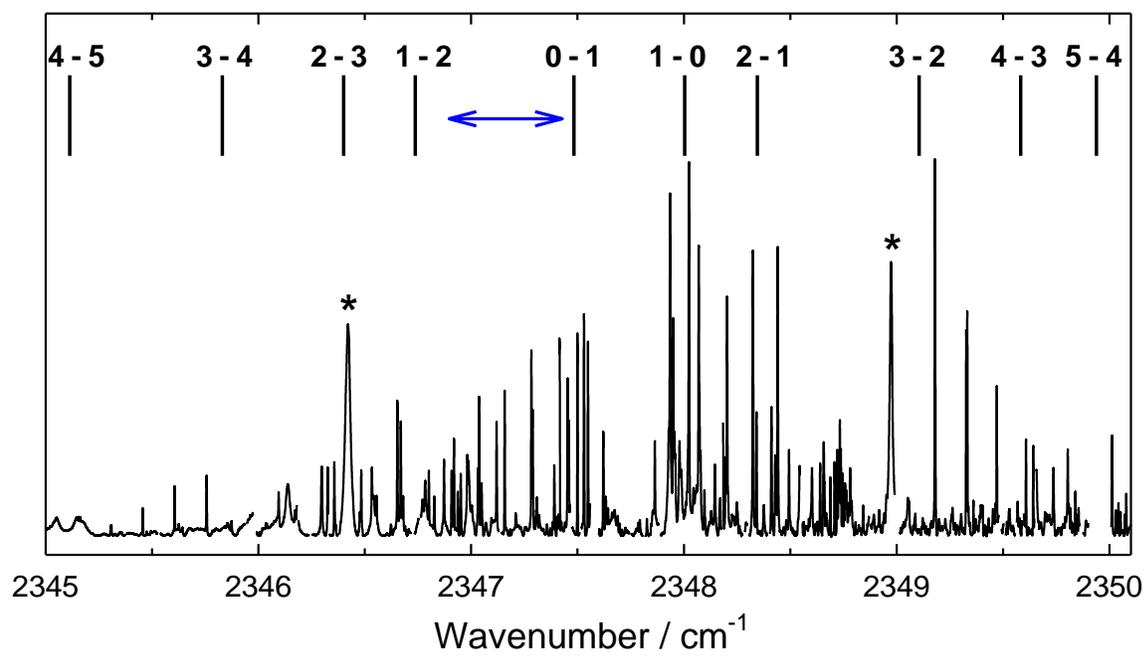

Fig. 3.

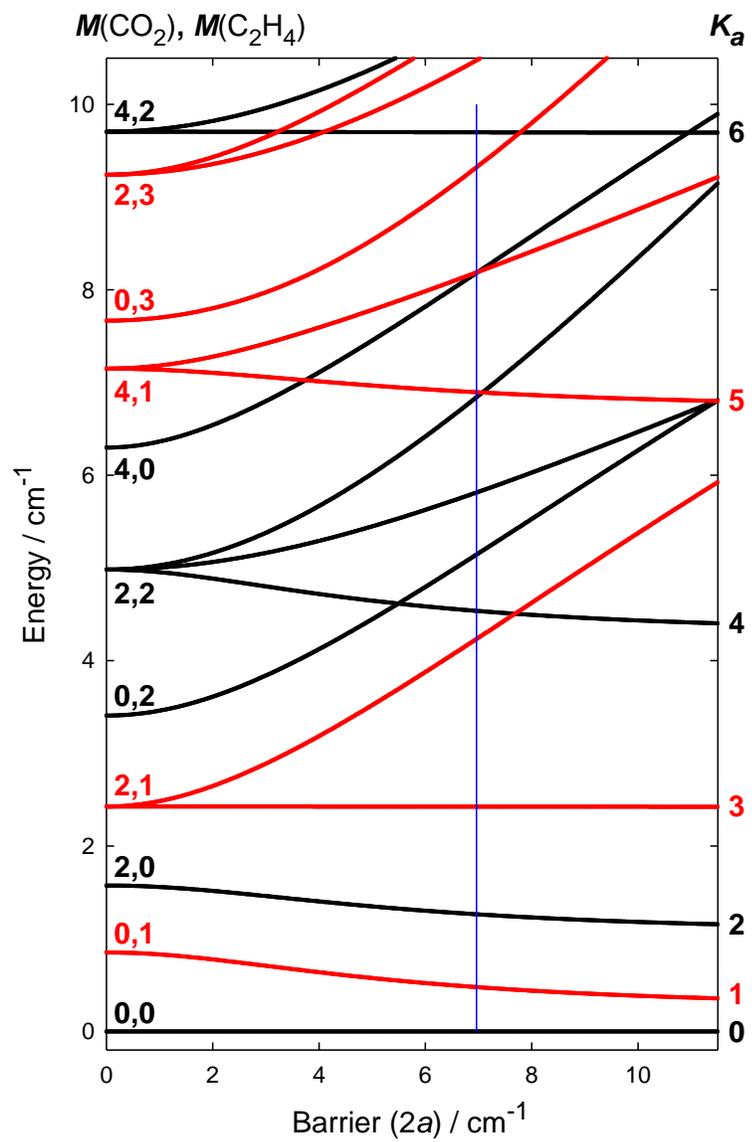

Fig. 4

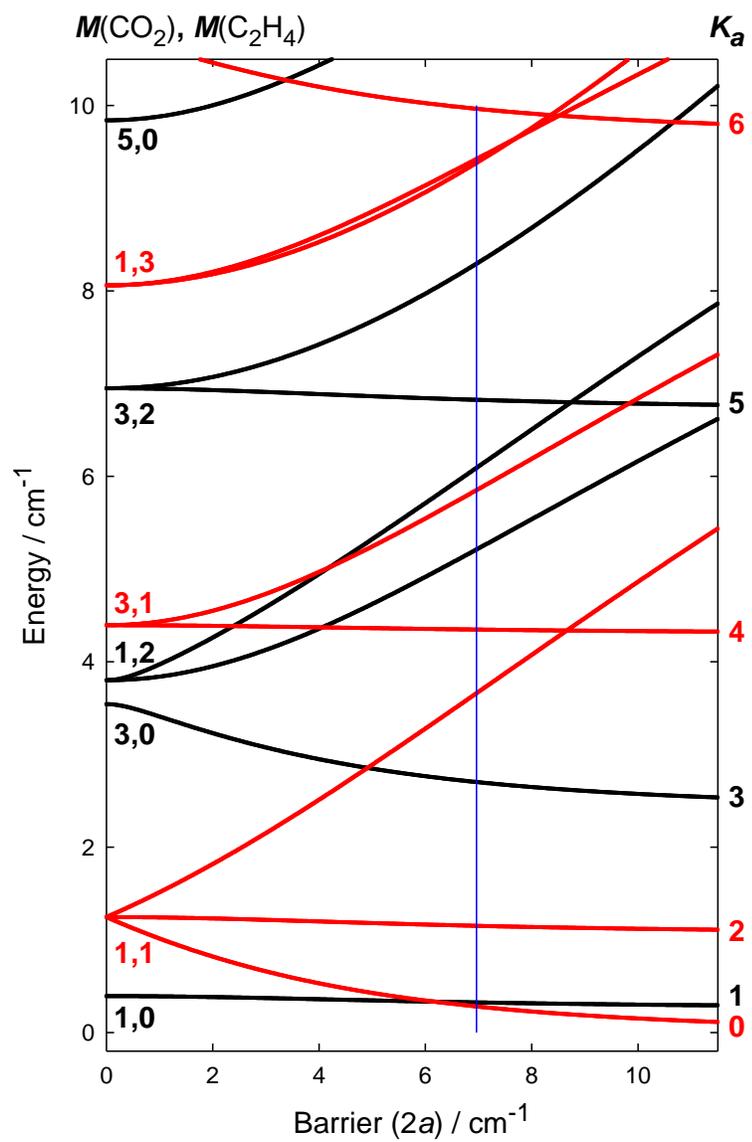

Fig. 5

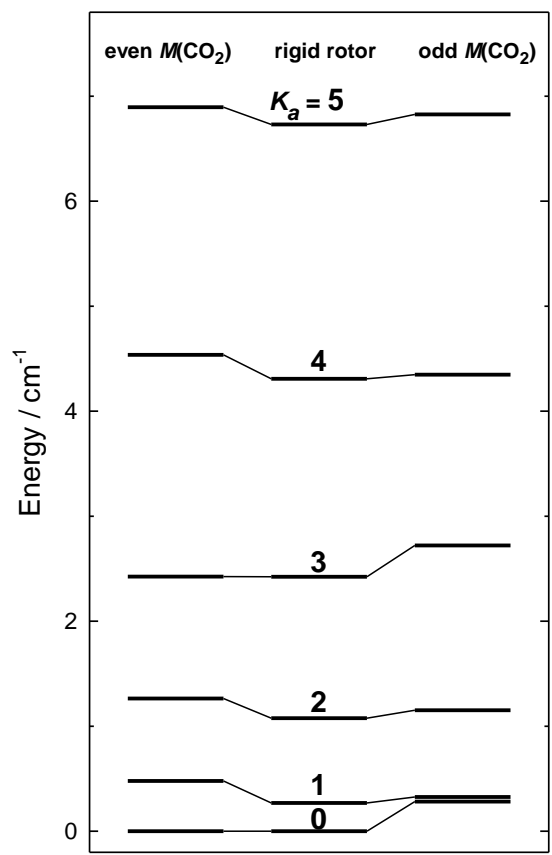

Fig. 6

Supplementary Information for:
The ethylene – carbon dioxide complex and the double rotor model

Table A-1. Assignments and observed and calculated line positions of transitions in the fundamental band of $CO_2$-$C_2H_4$ accompanying the $\nu_3$ band of $CO_2$ (in $cm^{-1}$).

Notes: Where $K_c$ is omitted, asymmetry doubling was not resolved and the transition was fitted to the mean of the two calculated transitions. The calculated line positions correspond to the individual experimental subband origins in Table I of the paper and the rotational parameters in Table II.

| $J'$ | $Ka'$ | $Kc'$ | $J''$ | $Ka''$ | $Kc''$ | Observed | Calculated | Obs - Calc |
|------|-------|-------|-------|--------|--------|----------|------------|------------|
| $K = 1 \square 0$ | | | | | | | | |
| 1 | 1 | 0 | 1 | 0 | 1 | 2347.9348 | 2347.9369 | -0.0021 |
| 2 | 1 | 1 | 2 | 0 | 2 | 2347.9495 | 2347.9536 | -0.0041 |
| 3 | 1 | 2 | 3 | 0 | 3 | 2347.9792 | 2347.9803 | -0.0011 |
| 4 | 1 | 3 | 4 | 0 | 4 | 2348.0235 | 2348.0188 | 0.0047 |
| 5 | 1 | 4 | 5 | 0 | 5 | 2348.0694 | 2348.0711 | -0.0017 |
| 1 | 1 | 1 | 0 | 0 | 0 | 2348.0694 | 2348.0697 | -0.0003 |
| 2 | 1 | 2 | 1 | 0 | 1 | 2348.2027 | 2348.2015 | 0.0012 |
| 3 | 1 | 3 | 2 | 0 | 2 | 2348.3241 | 2348.3246 | -0.0005 |
| 4 | 1 | 4 | 3 | 0 | 3 | 2348.4405 | 2348.4403 | 0.0002 |
| 5 | 1 | 5 | 4 | 0 | 4 | 2348.5510 | 2348.5503 | 0.0007 |
| 6 | 1 | 6 | 5 | 0 | 5 | 2348.6567 | 2348.6570 | -0.0003 |
| 7 | 1 | 7 | 6 | 0 | 6 | 2348.7627 | 2348.7627 | 0.0001 |
| 1 | 1 | 1 | 2 | 0 | 2 | 2347.6212 | 2347.6216 | -0.0004 |
| 2 | 1 | 2 | 3 | 0 | 3 | 2347.4587 | 2347.4577 | 0.0010 |
| 3 | 1 | 3 | 4 | 0 | 4 | 2347.2889 | 2347.2895 | -0.0006 |
| 4 | 1 | 4 | 5 | 0 | 5 | 2347.1196 | 2347.1197 | -0.0001 |
| 5 | 1 | 5 | 6 | 0 | 6 | 2346.9512 | 2346.9507 | 0.0005 |
| 6 | 1 | 6 | 7 | 0 | 7 | 2346.7841 | 2346.7847 | -0.0006 |
| 7 | 1 | 7 | 8 | 0 | 8 | 2346.6235 | 2346.6230 | 0.0005 |
| $K = 0 \square 1$ | | | | | | | | |
| 1 | 0 | 1 | 1 | 1 | 0 | 2347.5491 | 2347.5488 | 0.0003 |
| 2 | 0 | 2 | 2 | 1 | 1 | 2347.5303 | 2347.5303 | 0.0000 |
| 3 | 0 | 3 | 3 | 1 | 2 | 2347.4997 | 2347.4996 | 0.0001 |
| 4 | 0 | 4 | 4 | 1 | 3 | 2347.4537 | 2347.4540 | -0.0003 |
| 5 | 0 | 5 | 5 | 1 | 4 | 2347.3908 | 2347.3909 | -0.0001 |
| 6 | 0 | 6 | 6 | 1 | 5 | 2347.3092 | 2347.3090 | 0.0002 |
| 0 | 0 | 0 | 1 | 1 | 1 | 2347.4151 | 2347.4163 | -0.0012 |
| 1 | 0 | 1 | 2 | 1 | 2 | 2347.2831 | 2347.2833 | -0.0002 |
| 2 | 0 | 2 | 3 | 1 | 3 | 2347.1575 | 2347.1577 | -0.0002 |
| 3 | 0 | 3 | 4 | 1 | 4 | 2347.0374 | 2347.0376 | -0.0002 |
| 4 | 0 | 4 | 5 | 1 | 5 | 2346.9195 | 2346.9199 | -0.0004 |
| 5 | 0 | 5 | 6 | 1 | 6 | 2346.8014 | 2346.8016 | -0.0002 |
| 6 | 0 | 6 | 7 | 1 | 7 | 2346.6805 | 2346.6803 | 0.0003 |
| 7 | 0 | 7 | 8 | 1 | 8 | 2346.5554 | 2346.5551 | 0.0003 |
| 2 | 0 | 2 | 1 | 1 | 1 | 2347.8633 | 2347.8627 | 0.0006 |

| | | | | | | Obs. | Calc. | Diff. |
|---|---|---|---|---|---|---|---|---|
| 3 | 0 | 3 | 2 | 1 | 2 | | 2348.0225 | |
| 4 | 0 | 4 | 3 | 1 | 3 | 2348.1832 | 2348.1829 | 0.0003 |
| 5 | 0 | 5 | 4 | 1 | 4 | 2348.3409 | 2348.3408 | 0.0001 |
| 6 | 0 | 6 | 5 | 1 | 5 | 2348.4937 | 2348.4936 | 0.0001 |
| 7 | 0 | 7 | 6 | 1 | 6 | 2348.6400 | 2348.6406 | -0.0006 |

$K = 2 \leftarrow 1$

| | | | | | | | | |
|---|---|---|---|---|---|---|---|---|
| 2 | 2 | 0 | 2 | 1 | 1 | 2348.0962 | 2348.0967 | -0.0005 |
| 3 | 2 | 1 | 3 | 1 | 2 | 2348.0777 | 2348.0780 | -0.0003 |
| 4 | 2 | 2 | 4 | 1 | 3 | | 2348.0600 | |
| 5 | 2 | 3 | 5 | 1 | 4 | | 2348.0475 | |
| 2 | 2 | 1 | 2 | 1 | 2 | 2348.1452 | 2348.1454 | -0.0002 |
| 3 | 2 | 2 | 3 | 1 | 3 | 2348.1701 | 2348.1711 | -0.0010 |
| 4 | 2 | 3 | 4 | 1 | 4 | 2348.2072 | 2348.2058 | 0.0014 |
| 5 | 2 | 4 | 5 | 1 | 5 | 2348.2482 | 2348.2499 | -0.0017 |
| 2 | 2 | 1 | 1 | 1 | 0 | 2348.4111 | 2348.4110 | 0.0002 |
| 3 | 2 | 2 | 2 | 1 | 1 | 2348.5438 | 2348.5436 | 0.0002 |
| 4 | 2 | 3 | 3 | 1 | 2 | 2348.6682 | 2348.6678 | 0.0004 |
| 5 | 2 | 4 | 4 | 1 | 3 | | 2348.7840 | |
| 6 | 2 | 5 | 5 | 1 | 4 | | 2348.8927 | |
| 2 | 2 | 0 | 1 | 1 | 1 | 2348.4290 | 2348.4291 | -0.0001 |
| 3 | 2 | 1 | 2 | 1 | 2 | 2348.6015 | 2348.6009 | 0.0007 |
| 4 | 2 | 2 | 3 | 1 | 3 | 2348.7879 | 2348.7889 | -0.0010 |
| 5 | 2 | 3 | 4 | 1 | 4 | | 2348.9974 | |
| 6 | 2 | 4 | 5 | 1 | 5 | | 2349.2298 | |
| 2 | 2 | 1 | 3 | 1 | 2 | | 2347.6225 | |
| 3 | 2 | 2 | 4 | 1 | 3 | 2347.4414 | 2347.4422 | -0.0008 |
| 4 | 2 | 3 | 5 | 1 | 4 | 2347.2578 | 2347.2559 | 0.0019 |
| 5 | 2 | 4 | 6 | 1 | 5 | | 2347.0653 | |
| 2 | 2 | 0 | 3 | 1 | 3 | | 2347.7242 | |
| 3 | 2 | 1 | 4 | 1 | 4 | | 2347.6160 | |
| 4 | 2 | 2 | 5 | 1 | 5 | | 2347.5259 | |

$K = 1 \leftarrow 2$

| | | | | | | | | |
|---|---|---|---|---|---|---|---|---|
| 2 | 1 | 1 | 2 | 2 | 0 | | 2346.9854 | |
| 3 | 1 | 2 | 3 | 2 | 1 | | 2347.0051 | |
| 4 | 1 | 3 | 4 | 2 | 2 | | 2347.0271 | |
| 5 | 1 | 4 | 5 | 2 | 3 | | 2347.0475 | |
| 2 | 1 | 2 | 2 | 2 | 1 | 2346.9373 | 2346.9357 | 0.0016 |
| 3 | 1 | 3 | 3 | 2 | 2 | 2346.9084 | 2346.9087 | -0.0003 |
| 4 | 1 | 4 | 4 | 2 | 3 | 2346.8724 | 2346.8724 | 0.0000 |
| 5 | 1 | 5 | 5 | 2 | 4 | 2346.8268 | 2346.8269 | -0.0001 |
| 3 | 1 | 3 | 2 | 2 | 0 | 2347.3565 | 2347.3564 | 0.0001 |
| 4 | 1 | 4 | 3 | 2 | 1 | 2347.4653 | 2347.4652 | 0.0001 |
| 5 | 1 | 5 | 4 | 2 | 2 | 2347.5594 | 2347.5586 | 0.0008 |
| 6 | 1 | 6 | 5 | 2 | 3 | | 2347.6333 | |
| 3 | 1 | 2 | 2 | 2 | 1 | | 2347.4583 | |
| 4 | 1 | 3 | 3 | 2 | 2 | 2347.6434 | 2347.6379 | 0.0055 |
| 5 | 1 | 4 | 4 | 2 | 3 | | 2347.8239 | |
| 6 | 1 | 5 | 5 | 2 | 4 | | 2348.0153 | |
| 2 | 1 | 1 | 2 | 2 | 0 | 2346.6532 | 2346.6534 | -0.0002 |
| 2 | 1 | 2 | 3 | 2 | 1 | 2346.4837 | 2346.4825 | 0.0012 |
| 3 | 1 | 3 | 4 | 2 | 2 | 2346.2978 | 2346.2978 | 0.0000 |
| 4 | 1 | 4 | 5 | 2 | 3 | 2346.0956 | 2346.0960 | -0.0004 |
| 5 | 1 | 5 | 6 | 2 | 4 | 2345.8739 | 2345.8740 | -0.0001 |
| 1 | 1 | 0 | 2 | 2 | 1 | 2346.6692 | 2346.6711 | -0.0019 |
| 2 | 1 | 1 | 3 | 2 | 2 | 2346.5336 | 2346.5376 | -0.0040 |
| 3 | 1 | 2 | 4 | 2 | 3 | | 2346.4124 | |
| 4 | 1 | 3 | 5 | 2 | 4 | 2346.2978 | 2346.2953 | 0.0025 |
| 5 | 1 | 4 | 6 | 2 | 5 | 2346.1855 | 2346.1863 | -0.0008 |
| $K = 3 \leftarrow 2$ | | | | | | | | |
| 3 | 3 | 0 | 3 | 2 | 1 | 2348.7268 | 2348.7269 | -0.0001 |
| 3 | 3 | 1 | 3 | 2 | 2 | 2348.7325 | 2348.7314 | 0.0011 |
| 4 | 3 | 1 | 4 | 2 | 2 | 2348.7200 | 2348.7202 | -0.0002 |
| 4 | 3 | 2 | 4 | 2 | 3 | 2348.7325 | 2348.7336 | -0.0011 |
| 5 | 3 | 2 | 5 | 2 | 3 | 2348.7078 | 2348.7079 | -0.0001 |
| 5 | 3 | 3 | 5 | 2 | 4 | 2348.7375 | 2348.7381 | -0.0006 |
| 6 | 3 | 3 | 6 | 2 | 4 | 2348.6882 | 2348.6881 | 0.0001 |
| 6 | 3 | 4 | 6 | 2 | 5 | 2348.7451 | 2348.7460 | -0.0009 |
| 7 | 3 | 4 | 7 | 2 | 5 | | 2348.6603 | |
| 7 | 3 | 5 | 7 | 2 | 6 | | 2348.7586 | |
| 3 | 3 | | 2 | 2 | | 2349.1799 | 2349.1796 | 0.0003 |
| 4 | 3 | 1 | 3 | 2 | 2 | 2349.3312 | 2349.3311 | 0.0001 |
| 4 | 3 | 2 | 3 | 2 | 1 | 2349.3265 | 2349.3263 | 0.0002 |
| 5 | 3 | 2 | 4 | 2 | 3 | | 2349.4842 | |
| 5 | 3 | 3 | 4 | 2 | 2 | 2349.4702 | 2349.4698 | 0.0004 |
| 6 | 3 | 3 | 5 | 2 | 4 | 2349.6413 | 2349.6410 | 0.0003 |
| 6 | 3 | 4 | 5 | 2 | 3 | 2349.6075 | 2349.6072 | 0.0003 |
| 7 | 3 | 4 | 6 | 2 | 5 | 2349.8036 | 2349.8035 | 0.0001 |
| 7 | 3 | 5 | 6 | 2 | 4 | 2349.7364 | 2349.7362 | 0.0002 |
| $K = 2 \leftarrow 3$ | | | | | | | | |
| 3 | 2 | 1 | 3 | 3 | 0 | | 2346.7809 | |
| 3 | 2 | 2 | 3 | 3 | 1 | | 2346.7739 | |
| 4 | 2 | 2 | 4 | 3 | 1 | | 2346.7907 | |
| 4 | 2 | 3 | 4 | 3 | 2 | | 2346.7703 | |
| 5 | 2 | 3 | 5 | 3 | 2 | | 2346.8086 | |
| 5 | 2 | 4 | 5 | 3 | 3 | | 2346.7635 | |
| 2 | 2 | | 3 | 3 | | 2346.3261 | 2346.3262 | -0.0001 |
| 3 | 2 | 2 | 4 | 3 | 1 | 2346.1726 | 2346.1729 | -0.0003 |
| 3 | 2 | 1 | 4 | 3 | 2 | 2346.1809 | 2346.1804 | 0.0005 |
| 4 | 2 | 3 | 5 | 3 | 2 | | 2346.0170 | |
| 4 | 2 | 2 | 5 | 3 | 3 | 2346.0394 | 2346.0395 | -0.0001 |
| 5 | 2 | 4 | 6 | 3 | 3 | | 2345.8555 | |
| 5 | 2 | 3 | 6 | 3 | 4 | | 2345.9070 | |
| $K = 4 \leftarrow 3$ | | | | | | | | |
| 4 | 4 | | 4 | 3 | | | 2349.0556 | |
| 5 | 4 | 1 | 5 | 3 | 2 | | 2349.0506 | |
| 5 | 4 | 2 | 5 | 3 | 3 | | 2349.0521 | |

| | | | | | | Obs | Calc | Diff |
|---|---|---|---|---|---|---|---|---|
| 6 | 4 | 2 | 6 | 3 | 3 | | 2349.0438 | |
| 6 | 4 | 3 | 6 | 3 | 4 | | 2349.0485 | |
| 4 | 4 | | 3 | 3 | | 2349.6563 | 2349.6563 | 0.0000 |
| 5 | 4 | | 4 | 3 | | 2349.8036 | 2349.8036 | 0.0000 |
| 6 | 4 | 2 | 5 | 3 | 3 | 2349.9501 | 2349.9518 | |
| 6 | 4 | 3 | 5 | 3 | 2 | 2349.9501 | 2349.9501 | |
| $K = 3 \rightarrow 4$ | | | | | | | | |
| 4 | 3 | | 4 | 4 | | | 2346.3571 | 2346.3551 | 0.0019 |
| 5 | 3 | 2 | 5 | 4 | 1 | 2346.3571 | 2346.3580 | -0.0009 |
| 5 | 3 | 3 | 5 | 4 | 2 | 2346.3571 | 2346.3571 | 0.0000 |
| 6 | 3 | 3 | 6 | 4 | 2 | 2346.3571 | 2346.3608 | -0.0037 |
| 6 | 3 | 4 | 6 | 4 | 3 | 2346.3571 | 2346.3582 | -0.0011 |
| 3 | 3 | | 4 | 4 | | 2345.7569 | 2345.7555 | 0.0014 |
| 4 | 3 | | 5 | 4 | | 2345.6067 | 2345.6074 | -0.0007 |
| 5 | 3 | | 6 | 4 | | 2345.4570 | 2345.4584 | -0.0014 |
| 6 | 3 | 4 | 7 | 4 | 3 | 2345.3063 | 2345.3064 | -0.0001 |
| 6 | 3 | 3 | 7 | 4 | 4 | 2345.3094 | 2345.3093 | 0.0001 |
| 7 | 3 | 5 | 8 | 4 | 4 | 2345.1558 | 2345.1516 | 0.0042 |
| 7 | 3 | 4 | 8 | 4 | 5 | 2345.1641 | 2345.1589 | 0.0052 |
| $K = 5 \rightarrow 4$ | | | | | | | | |
| 5 | 5 | | 5 | 4 | | | 2349.2637 | |
| 6 | 5 | | 6 | 4 | | | 2349.2591 | |
| 5 | 5 | | 4 | 4 | | 2350.0118 | 2350.0114 | 0.0004 |
| 6 | 5 | | 5 | 4 | | 2350.1574 | 2350.1582 | -0.0008 |
| 7 | 5 | | 6 | 4 | | 2350.3085 | 2350.3048 | 0.0037 |
| $K = 4 \rightarrow 5$ | | | | | | | | |
| 4 | 4 | | 5 | 5 | | 2345.0381 | 2345.0379 | 0.0003 |
| 5 | 4 | | 6 | 5 | | 2344.8861 | 2344.8863 | -0.0002 |
| 6 | 4 | | 7 | 5 | | 2344.7343 | 2344.7345 | -0.0002 |
| 7 | 4 | | 8 | 5 | | 2344.5830 | 2344.5826 | 0.0004 |